# A Feasibility Study on Programmer Specific Instruction Set Processors

T.M.R.L.B. Abeysinghe, N. Hassan
Department of Statistics and Computer Science,
Faculty of Science, University of Peradeniya,
Peradeniya, Sri Lanka.
rashmie0045@gmail.com, naadiyahassan@gmail.com

R.G. Ragel
Department of Computer Engineering,
Faculty of Engineering, University of Peradeniya,
Peradeniya, Sri Lanka.
roshanr@pdn.ac.lk

*Abstract*—ASIPs are designed in order to execute instructions of a particular domain of applications. The designing of ASIPs addresses the major challenges faced by a system on chip such as size, cost, performance and energy consumption. The higher the number of similar instructions within the domain to be mapped the lesser the energy consumption, the smaller the size and the higher the performance of the ASIP. Thus, designing processors for domains with more similar programs would overcome these issues. This paper describes the investigation of whether the domains of programmer specific programs have any significance like application specific program domains and thus, whether the approach of designing processors known as Programmer Specific Instruction Set Processors is worthwhile. We performed the evaluation at the instruction level by using four different measures to obtain the similarity of programs: (1) by the existence of each instruction, (2) by the frequency of each instruction, (3) by two consecutive instruction patterns and (4) by three consecutive instruction patterns of application specific and programmer specific programs. We found that although programmer specific instructions show some impact on the similarity measures, they are much smaller and therefore insignificant compared to the impact from application specific programs.

## I. Introduction

Processors are designed to be able to execute a set of instructions. Hence, the architecture of a processor depends on the type and the size of the instruction set that the processor is intended to execute. The larger the size of the instruction set, the higher the complexity of the processor will be.

General Purpose Processors (GPPs) are processors, which can be used for multiple applications. Therefore, by considering all the applications that are intended to be run on the processor, all possible instructions are hardwired on them. However, having such a large instruction set leads to lower performance, higher energy consumption and higher dimensions in a processor.

With the extensive usage of processors for various kinds of devices apart from computers, such as mobile phones, cars, cameras, etc., the need for more efficient processors started to arise. Therefore, it is clear that all the processors are not intended to be used for similar kinds of applications.

One of the ways of achieving such efficient processors was accomplished by using Application Specific Instruction set Processors (ASIPs) [1]. ASIPs are designed targeting specific domains of applications. Only a set of selected instructions are hardwired on them according to the domain of applications that the ASIP will be used for. Thus, ASIPs are attributed for high performance, reduced size and lower energy consumption compared to those of GPPs. Although ASIPs have many advantages over GPPs, the process of selecting instructions to be mapped on to them is an additional task aggregated to its design process. Since complex application programs use hundreds of types of processor instructions, selecting the most suitable instructions in order to maximize the performance in an optimized way is a major challenge in the designing process of an ASIP. Thus, the designing process of ASIPs is mainly concerned on instruction set generation/selection for a given application domain. If the similarity between the application programs within the domain is higher, then the instruction set to be mapped becomes smaller reducing the complexity of the ASIP and its cost. Therefore, if there is any other way of grouping the programs (other than grouping them as application specific domains), which would have more similarities between them, it would be useful for building processors for such groups of applications.

Programming is becoming easier and easier with the set of design tools and domain specific languages that are being introduced to developers. Therefore, in the future, it could be expected that competent users will also become programmers regardless of their profession. Users becoming programmers and programmable logic devices (such as FPGAs) would bring the possibility of building processors targeting such users. We define such processors (processors build targeting specific users) Programmer Specific Instruction Set Processors (PSISP).

The unique way, a programmer writes his programs or his style of coding could differ by the profession of the programmer, the company he works for and might even depend on factors like the region he lives, the age of the programmer and many more. Therefore, if this coding style is found to be persisting in each occasion, a PSISP could be developed. Hence, the concept of PSISPs discussed here is versatile to be applied in many situations.

For example, think of a future where there will be mobile phones with PSISPs targeting engineers, scientists and doctors. Will such different people be writing applications that are best run on special processors designed for them? This is a situation we are trying to study in this paper.

By taking the aforementioned scenarios into consideration, it is desirable that there is a possibility of building "Programmer Specific Instruction Set Processors". In this paper, we have performed a feasibility study on the aforementioned hypothesis by performing an instruction level study of programmer specific programs and application specific programs by using several different metrics: (1) Jaccard similarity (to measure similarity of programs considering the existence of instructions), (2) Cosine similarity (to measure similarity of programs considering the frequency of instructions) and (3) Euclidean distance (to measure similarity of programs considering two and three consecutive instruction patterns). This paper provides a comparison between the similarity measures of the above mentioned groups.

The organization of the rest of the paper is as follows: Section II provides details about the related research work that have been carried out in the past in this area and the methodologies that have been used. A description about the methodology that has been followed to achieve the outcomes of this research is described in Section III. Section IV contains summarized results of the research while the Section V contains a discussion about the results. Finally, the Section VI provides the conclusion.

## II. RELATED WORK

ASIPs have a history of more than two decades. Most of the related researches concentrate on design issues of ASIPs and their instruction set generation and evaluation. Several methodologies have been proposed for instruction set generation of ASIPs such as the one presented by Alomary et al. in 1993 [1]. This paper concentrates on maximizing the performance of the chip under the constraints of chip area and its energy consumption with an optimized way of instruction selection. This method also enables designers to predict the performance of the chip before implementing their design.

According to Jain et al. [2], there are five steps that have been identified in the ASIP design process. They are, application analysis, architectural design space exploration, instruction set generation, code synthesis and hardware synthesis. Jain et al. have surveyed the status of this area and have identified some issues that need to be addressed. According to this survey the performance estimation of ASIPs is based either on scheduler based or simulation based techniques. Instruction set is generated either through synthesis [3], [4] or a selection [1] process. Code is synthesized either by a re-targetable code generator [5] or by a custom generated compiler.

Instruction set generation and evaluation is a popular area of research in ASIPs. Jason Cong [10] has addressed the problem of generating application specific instructions to improve the execution speed for configurable processors. Pattern generation, pattern selection and application mapping algorithms have been proposed to efficiently utilize the extensibility of the target configurable processor.

Jääskelainen [11] has described about constructing a tool that assists in co-designing application specific instruction set-processors for embedded systems. The toolset is based on a customizable processor architecture template, which is VLIW-derived architecture paradigm called Transport Triggered Architecture (TTA). The toolset has addressed some of the pressing shortcomings in the existing toolsets, such as automated exploration of the design space, limited runtime re-target-ability of the design tools or restrictions in the customization of the target processors.

The extension of a given instruction set with specialized instructions has become a common technique used to speed up the execution of applications. Galuzzi and Bertels [12] presents a thorough analysis of the issues involved during the customization of an instruction set by means of a set of specialized application specific instructions.

To our knowledge, no work has been undertaken to utilize the uniqueness of a programmer in the way he writes his codes, to build processors. Therefore, this is the first of such work and in this research we have investigated whether using programmers' uniqueness in programming to build processors is worthwhile. Our study only considers instruction selection and therefore instruction synthesis is left as future work.

## III. METHODOLOGY

As described above the aim of this research is to investigate whether the idea of designing PSISPs is worthwhile by means of measuring and comparing the similarities of the instruction sets of a collection of programs grouped according to the programmers (named as programmer specific programs), according to the application (named as application specific) and a grouping where the programs are entirely different from each other (named as totally different).

Since ASIPs are an accepted concept, we have used them as a standard by which the appropriateness of PSISPs could be described. We expected that the similarity between Application Specific Programs would be much higher than that of Totally Different Programs. Just like that if the similarity between Programmer Specific Programs vastly deviates from that of Totally Different Programs and gets closer to that of Application Specific Programs, then the hypothesis discussed in this paper would be proved. Firstly, the datasets had to be grouped as Application Specific, Programmer Specific and Totally Different and the similarity values according to several metrics were obtained. How the grouping was done is described in subsection A and the similarity measures used have been described in subsection B.

### A. Datasets used and the grouping of each dataset

In order to conduct this experiment six data sets (which were programs written in C programming language) were collected. The data sets 1-5 were 5x5 sets which contained programs written by the students of a programming class. Each 5x5 set consisted of 25 programs which were written by five programmers for five applications. Grouping was as follows.

Fig. 1 shows an Application Specific Program subset.

|  | Programmer 1 | Programmer 2 | Programmer 3 | Programmer 4 | Programmer 5 |
|---|---|---|---|---|---|
| Application 1 | Prg1App1 | Prg2App1 | Prg3App1 | Prg4App1 | Prg5App1 |
| Application 2 | Prg1App2 | Prg2App2 | Prg3App2 | Prg4App2 | Prg5App2 |
| Application 3 | Prg1App3 | Prg2App3 | Prg3App3 | Prg4App3 | Prg5App3 |
| Application 4 | **Prg1App4** | **Prg2App4** | **Prg3App4** | **Prg4App4** | **Prg5App4** |
| Application 5 | Prg1App5 | Prg2App5 | Prg3App5 | Prg4App5 | Prg5App5 |

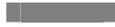 An Application Specific subset

Fig. 1. An Application Specific Subset of a 5x5 set

a) Application Specific Programs – Data/programs were grouped according to the programmers in order to have 5 subsets each consisting of 5 programs written by the same programmer. Note that, all the programs of this subset are written for the same application. There are five such subsets for each 5x5 set in this grouping.

b) Programmer Specific Programs - Fig. 2 gives an example for a Programmer Specific Programs subset of a 5x5 set. Data/programs were grouped according to the applications to which the program was written for (having 5 subsets each consisting of 5 programs written for the same application). Note that all the programs of this subset are written by the same programmer. There are five such subsets for each 5x5 set in this grouping.

c) Totally Different Programs grouping 1

d) Totally Different Programs grouping 2

e) Totally Different Programs grouping 3

Totally Different Programs are neither written by the same programmer nor written for the same application, hence considered to be entirely different from each other (having 5 subsets each consisting of 5 totally different programs). Fig. 3 shows an example of a subset of Totally Different Programs.

|  | Programmer 1 | Programmer 2 | Programmer 3 | Programmer 4 | Programmer 5 |
|---|---|---|---|---|---|
| Application 1 | Prg1App1 | **Prg2App1** | Prg3App1 | Prg4App1 | Prg5App1 |
| Application 2 | Prg1App2 | **Prg2App2** | Prg3App2 | Prg4App2 | Prg5App2 |
| Application 3 | Prg1App3 | **Prg2App3** | Prg3App3 | Prg4App3 | Prg5App3 |
| Application 4 | Prg1App4 | **Prg2App4** | Prg3App4 | Prg4App4 | Prg5App4 |
| Application 5 | Prg1App5 | **Prg2App5** | Prg3App5 | Prg4App5 | Prg5App5 |

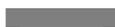 A programmer Specific subset

Fig. 2. A programmer Specific subset of a 5x5 set

|  | Programmer 1 | Programmer 2 | Programmer 3 | Programmer 4 | Programmer 5 |
|---|---|---|---|---|---|
| Application 1 | **Prg1App1** | Prg2App1 | Prg3App1 | Prg4App1 | Prg5App1 |
| Application 2 | Prg1App2 | Prg2App2 | Prg3App2 | **Prg4App2** | Prg5App2 |
| Application 3 | Prg1App3 | **Prg2App3** | Prg3App3 | Prg4App3 | Prg5App3 |
| Application 4 | Prg1App4 | Prg2App4 | Prg3App4 | Prg4App4 | **Prg5App4** |
| Application 5 | Prg1App5 | Prg2App5 | **Prg3App5** | Prg4App5 | Prg5App5 |

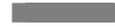 A Totally Different Group

Fig. 3. A Totally Different Subset of the 5x5 set

Note that, this subset contains programs written by different programmers for different applications. There are five such subsets for each 5x5 set in this grouping.

We expected the Totally Different Programs to be entirely different from each other and since the similarity value could differ with the way they have been grouped in to subsets, we have taken three different groupings for Totally Different Programs and we have taken the average similarity value of those 3 as the similarity value of Totally Different Programs.

The dataset 6 was a 3x3 set, which consisted of programs written by experienced programmers. This dataset contained nine programs which were written for three different applications by three different programmers. The grouping of dataset 6 was similar to the grouping of other datasets.

1) Application Specific Programs - three subsets each consisting of three programs written by the same programmer.

2) Programmer Specific Programs – three subsets each consisting of three programs written for the same application.

3) Totally Different Programs grouping 1

4) Totally Different Programs grouping 2

The two Totally Different Programs grouping consisted of three subsets each consisting of three programs which were entirely different from each other.

To retrieve the instructions used, each program was compiled to its ARM Thumb assembly code using the GCC-cross compiler. Then the similarity of programs was measured in four ways as mentioned below in subsection B.

B. Similarity Measures

*1) Considering the existence of each ARM-Thumb instruction in the program*

All the assembly instructions of each assembly code were extracted regardless of their number of occurrences and each mnemonic of the instructions was considered as a different instruction. Extracted assembly instructions of each program were then recorded using 1's and 0's to represent the presence and absence of all the possible ARM-thumb instructions in each and every program respectively. Jaccard similarity was used to measure the similarity of programs.

Jaccard similarity [9] between two sets of words S1 and S2 (two documents represented as vectors/sets having frequencies of the appearance of each and every word in the document) is given by,

$$JaccardSimilarity(S1, S2) = (S1 \cap S2)/(S1 \cup S2) \quad (1)$$

Then considering each data set grouped as Application Specific, Programmer Specific and Totally Different, Jaccard similarity values were calculated as follows.

First all possible combinations of two program sets within a subset was identified and Jaccard similarity measures were obtained for each of those combinations of two programs. Then the average similarity value of each subset and thus the average similarity of all the subsets of a group were obtained.

Then these average Jaccard similarity measures of application specific data, programmer specific data and totally different data were normalized and compared with each other.

*2) By the frequency of occurrence of each instruction*

The number of times an instruction is used in a program differs from program to program. Two programs having similar number of frequencies for their instructions could be considered as having a higher degree of similarity than two programs with different number of frequencies for their instructions. Therefore this frequency of occurrence of each instruction could be used to measure the similarity between two programs and could be taken as a more accurate method of measuring similarity of two programs than just considering the existence of instructions.

For the purpose of measuring similarity, the bag-of-words model was used [14]. Bag-of-words model is used to classify documents only by considering the frequency of occurrence of words, but not the order in which they appear.

In text domains, a document is generally treated as a bag-of-words, where each unique word in the vocabulary is a dimension of a vector. Thus, the similarity between two documents could be found by finding the cosine similarity between the vectors corresponding to these two documents [14]. This cosine value would take a higher value for two programs having a higher degree of similarity and consequently this would be lower if the two programs have lower similarity. Hence, this concept was used to measure the similarity between two programs by means of frequency of occurrence of instructions. Equation (2) was used to get the cosine similarity value between two programs. In (2), A and B are the frequency vectors of the two documents and θ is the angle between them.

$$similarity = \cos\theta = \frac{A \cdot B}{\|A\|\|B\|}$$

$$= \frac{\sum_{i=1}^{n} A_i \times B_i}{\sqrt{\sum_{i=1}^{n}(A_i)^2} \times \sqrt{\sum_{i=1}^{n}(B_i)^2}} \quad (2)$$

Where,

$A_i$ – frequency of instruction 'i' in program A

$B_i$ – frequency of instruction 'i' in program B

$\cos\theta$ – cosine similarity value

Here the angular distance (cosine) between two vectors is measured.

To start with, the frequencies of each instruction were extracted from its assembly file.

As mentioned earlier, each grouping of a dataset (Application Specific, Programmer Specific and Totally Different) has 5 subsets within them and the average similarity value of these 5 subsets were considered as the similarity value for that particular grouping. Since the cosine value gives the similarity value between two programs, to obtain the similarity value within a subset all the possible combinations of two programs within that subset had to be considered. Average of similarity values of all the possible combinations of two programs was considered as the average similarity value of that subset. In this manner, the cosine similarity value of Application Specific, Programmer Specific and Totally Different groupings were obtained.

*3) By two consequent instructions patterns*

Frequency of occurrence of each instruction might be an accurate method to measure similarity between two programs than just considering the existence of each instruction. But, it might be difficult to get an accurate value for similarity of two programs just by considering this. The order or the flow of instructions is a really important factor when comparing programs. Two programs might be quite similar in terms of their instruction frequencies, but their flow of instructions might be totally different from each other. Thus, patterns of instructions also had to be considered in order to get a more accurate value for similarity. Each pair of consequent instructions was considered as a pattern and all such instruction patterns of a program were extracted from the assembly file of the program. When extracting patterns the control flow of the program also had to be considered (using basic blocks, branch instructions, etc.)

Once all the existing instruction patterns of the 5x5 set were found, the presence of each pattern was listed against each program in a Boolean data table. From this Boolean representation of programs' characteristics we produced a vector representation that can be used to calculate the distance between two programs [13].

If each instruction pattern available is considered as an independent variable, each program can be represented as a multi-variable equation P such that,

$$P = Aa + Bb + Cc + ... = (A, B, C, ...) \quad (3)$$

where, a, b, c… each is an instruction pattern and A, B, C… are coefficients, which are either 1 or 0 depending on the availability of each instruction pattern.

These vectors can be considered as coordinates in a multi-dimensional space. The distance between 2 such points is proportional to the difference between the corresponding programs. This distance between two points is assessed by calculating their Euclidean distance. Therefore the following equation was used to calculate the distance between two programs.

If each one of the points P1 and P2 can be represented as n independent variables, the Euclidean distance of P1 and P2, denoted by $D_{P1P2}$, is given in (4)

$$D_{P1P2} = \sqrt{(A_1-A_2)^2 + (B_1-B_2)^2 + \ldots + (n_1-n_2)^2} \quad (4)$$

Where,

A1, B1, …, n1 – coefficients of P1

A2, B2, …, n2 – coefficients of P2

Similar to the above methods, to obtain the similarity value within a subset, all the possible combinations of two programs within that subset had to be considered, because the distance value gives the similarity value between two programs. Therefore, in the same way, the distance similarity values of Application Specific, Programmer Specific and Totally Different groupings were obtained.

*4) By three consequent instruction patterns*

The flow of instructions would be represented better by using a higher number of consequent instructions as a pattern. Hence, it was decided to measure the similarity by considering three consequent instructions as patterns as well. Similar to the previous case, three consequent instruction patterns were extracted from the assembly code of each program. The procedure, which was followed to get the distance values was the same as the one used earlier.

IV. RESULTS

Table I shows the summary of the results obtained considering the existence of each instruction. Table II shows the summary of the results obtained by considering the frequency of instructions.

TABLE I. SUMMARY RESULTS CONSIDERING EXISTENCE OF INSTRUCTIONS.

| Data Set | Programmer Specific | Application Specific | Totally Different 1 | Totally Different 2 | Totally Different 3 |
|---|---|---|---|---|---|
| 1 | 0.4453 | 0.6697 | 0.4557 | 0.4465 | 0.4500 |
| 2 | 0.5075 | 0.6699 | 0.5057 | 0.5003 | 0.5055 |
| 3 | 0.4594 | 0.6829 | 0.4613 | 0.4566 | 0.4655 |
| 4 | 0.4554 | 0.6816 | 0.4617 | 0.4493 | 0.4638 |
| 5 | 0.4971 | 0.6623 | 0.4885 | 0.4889 | 0.4815 |
| Average | 0.4729 | 0.6733 | 0.4720 | | |
| Normalized | 1.002 | 1.426 | 1 | | |
| 6 | 0.4879 | 0.6592 | 0.4932 | | |
| Normalized | 0.989 | 1.337 | 1 | | |

TABLE II. SUMMARY OF THE RESULTS CONSIDERING FREQUENCY OF INSTRUCTIONS.

| Data Set | Programmer Specific | Application Specific | Totally different 1 | Totally different 2 | Totally different 3 |
|---|---|---|---|---|---|
| 1 | 0.797 | 0.87 | 0.798 | 0.784 | 0.791 |
| 2 | 0.776 | 0.827 | 0.755 | 0.755 | 0.764 |
| 3 | 0.778 | 0.879 | 0.776 | 0.776 | 0.77 |
| 4 | 0.812 | 0.916 | 0.815 | 0.806 | 0.81 |
| 5 | 0.773 | 0.852 | 0.785 | 0.793 | 0.777 |
| Average | 0.787 | 0.869 | 0.784 | | |
| Normalized | 1.005 | 1.109 | 1 | | |
| 6 | 0.691 | 0.764 | 0.648 | | |
| Normalized | 1.066 | 1.179 | 1 | | |

V. DISCUSSION

According to the results of each table, it can be seen that the similarity values have not changed significantly with the dataset and therefore it is clear that the results do not depend on the datasets chosen.

The reason behind taking three Totally Different Groupings was to test whether the similarity value changes with the way we group them as Totally Different Programs and if so to minimize the effect of that by taking an average value. However, the three Totally Different Groups have proved otherwise by producing very close similarity values.

By considering Table I, Table II, Table III and Table IV, it is clear that the similarity between Programmer Specific Programs is lesser than the similarity between Application Specific Programs and it is much closer to Totally Different Programs when compared with all the four metrics.

Table III shows the summary of the results obtained by considering two consequent instruction patterns. Table IV shows the summary of the results obtained by considering three consequent instruction patterns.

TABLE III. SUMMARY RESULTS CONSIDERING TWO CONSEQUENT INSTRUCTION PATTERNS.

| Data Set | Programmer Specific | Application Specific | Totally different 1 | Totally different 2 | Totally different 3 |
|---|---|---|---|---|---|
| 1 | 7.95 | 6.43 | 8.02 | 8.09 | 8.07 |
| 2 | 8.19 | 6.8 | 8.24 | 8.3 | 8.22 |
| 3 | 7.89 | 6.03 | 7.89 | 7.96 | 7.97 |
| 4 | 8.23 | 6.27 | 8.2 | 8.27 | 8.19 |
| 5 | 8.22 | 6.65 | 8.31 | 8.2 | 8.3 |
| Average | 8.1 | 6.44 | 8.15 | | |
| Normalized | 1.007 | 1.266 | 1 | | |
| 6 | 9.39 | 8.53 | 9.44 | | |
| Normalized | 1.005 | 1.107 | 1 | | |

TABLE IV.  SUMMARY RESULTS CONSIDERING THREE CONSEQUENT INSTRUCTION PATTERNS.

| Data Set | Programmer Specific | Application Specific | Totally different 1 | Totally different 2 | Totally different 3 |
|---|---|---|---|---|---|
| 1 | 8.59 | 7.32 | 8.67 | 8.73 | 8.68 |
| 2 | 9.2 | 8.1 | 9.27 | 9.28 | 9.24 |
| 3 | 8.8 | 7.34 | 8.81 | 8.87 | 8.85 |
| 4 | 9.26 | 7.54 | 9.28 | 9.29 | 9.3 |
| 5 | 9.25 | 8.04 | 9.36 | 9.31 | 9.39 |
| Average | 9.02 | 7.67 | 9.09 | | |
| Normalized | 1.007 | 1.185 | 1 | | |
| 6 | 11.01 | 10.28 | 11.11 | | |
| Normalized | 1.009 | 1.081 | 1 | | |

However, considering all the results obtained in this research, the following conclusions can be made.

1. Since programs written by undergraduate students in a programming class were used in datasets 1-5, the following causes might have contributed to the results of those datasets.

   - There is a tendency that students work together in groups and therefore, such groups of students might get adapted to similar writing styles.

   - Since all of them had been taught C programming by the same lecturer and might have used the same text books as recommended by the lecturer they might all have certain similar programming styles.

   The programs written by these students were not that complicated and there might not have been many ways to write the same program. In other words complicated applications could be written in many ways to achieve the same task. Such a program would not be written the same way by several programmers. These factors might have contributed for these programmers to write similar programs for different applications and concealed their uniqueness in writing code. This might be a reason for Application Specific Programs to take a significant lead over Programmer Specific Programs in similarity in the datasets 1-5.

   Therefore, it was needed to check whether the source of the dataset had an impact on the results so the dataset 6 had to be introduced. Dataset 6 consisted of programs written by experienced programmers. Still the result was similar to the other datasets. However it is still not possible to say that the source of the dataset has not impacted the results since the dataset 6 too has not covered a broad spectrum of programmers/programs.

   Therefore, as future work the authors suggest an experiment to get data, which protects and enhances the uniqueness of programmers. This should be conducted with experienced programmers from different backgrounds and they should be asked to write several complicated programs.

2. The metrics we have used for the purpose of measuring similarity, are correct, since the results of application specific programs and totally different programs has a significant difference (ASIPs would not exist if such a difference is not there).

VI. CONCLUSION

In this paper, we have investigated whether the concept of Programmer Specific Instruction Set Processors is feasible. We have performed an instruction level study of Programmer Specific Programs and Application Specific Programs by using several different similarity metrics to prove our hypothesis of "Building Programmer Specific Instruction set Processors is worthwhile as the similarity between programmer specific programs is higher than or as good as that of application specific programs". Considering all the results obtained, it can be concluded that the results do not support the hypothesis that we have considered for our study.

REFERENCES

[1] A. Alomary, T. Nakata, Y. Honma, M. Imai, N. Hikichi, , "An ASIP instruction set optimization algorithm with functional module sharing constraint.", Proc. ICCAD-93, 7-11 Nov. 1993, pp. 526-532.

[2] M. K. Jain, M. Balakrishnan, and A. Kumar, "ASIP design methodologies: Survey and issues." in Proceedings of the IEEE / ACM International Conference on VLSI Design. (VLSI 2001), 2001, pp. 76–81.

[3] J. Huang and A. M. Despain, "Generating Instruction Sets and Microarchitectures from Applications.",IEEE Transactions on CAD of ICs and Systems, Vol. 14, No. 6, June, 1995.

[4] M. Gschwind, "Instruction set selection for ASIP design.", Proc. CODES-99, 3-5 May 1999, pp. 7-1 1.

[5] R. Leulpers, P. Marwedel, "Instruction set modelling for ASIP code generation" inproceedings of the 9th International Conference on VLSI Design: VLSI in Mobile Communication,1996, p.77

[6] J. V. Praet, G. Gosssens, D. Lanner, H. D. Man, "Instruction Set Definition and Instruction Selection for ASIPs" inproceedings of the 7th international symposium on High-level synthesis,1994, pp.11-16

[7] S. M. Z. Eissen, B. Stein, "Intrinsic Plagiarism Detection" in Proceedings of the 28th European Conference on IR Research, ECIR 2006, pp. 565-569, 2006.

[8] P. Juola, "Authorship Attribution",Foundation and trends in information retrieval., vol. No.3,2006,pp. 233-334,

[9] A. Huang, "Similarity measures for text document clustering,"In Proc. of the Sixth New Zealand Computer Science Research Student Conference NZCSRSC, 2008, pp. 49—56.

[10] Jason Cong, Yiping Fan, Guoling Han, Zhiru Zhang, "Application-Specific Instruction Generation for Configurable Processor Architecture", in proceedings of the 2004 ACM/SIGDA 12th international symposium on Field programmable gate arrays, 2004, pp.183-189.

[11] Jääskeläinen, P., Guzma, V., Cilio, A., Takala, J., "Codesign toolset for application-specific instruction-set processors." In Proc. SPIE - multimedia on mobile devices. 05070X-1-10,2007.

[12] Carlo Galuzzi , Koen Bertels, "The Instruction-Set Extension Problem: A Survey", ACM Transactions on Reconfigurable Technology and Systems (TRETS), v.4 n.2,May 2011 , pp.1-28

[13] M.M.E Karunarathna, Yu-Chu Tian, C. Fidge, R. Hayward,"Algorithm Clustering for Multi-algorithm Processor Design", in proceedings of IEEE 31st International Conference onComputer Design (ICCD), 2013, pp.451-454.

[14] Wikipedia,(2013, August). Cosine Similarity [Online]
Available: http://en.wikipedia.org/wiki/Cosine_similarity